\documentclass[12pt]{article}

\usepackage{graphics,epsfig}
\def\bb#1{\hbox{\mybb#1}}

\def\dalemb#1#2{{\vbox{\hrule height .#2pt
       \hbox{\vrule width.#2pt height#1pt \kern#1pt
               \vrule width.#2pt}
       \hrule height.#2pt}}}

\let\a=\alpha \let\b=\beta \let\g=\gamma \let\d=\delta \let\e=\epsilon
\let\z=\zeta  \let\th=\theta  \let\k=\kappa
\let\l=\lambda \let\m=\mu \let\n=\nu \let\x=\xi \let\p=\pi 
\let\s=\sigma \let\t=\tau    
\let\vp=\varphi \let\vep=\varepsilon
\let\w=\omega       \let\D=\Delta \let\Th=\Theta \let\L=\Lambda
 \let\P=\Pi \let\S=\Sigma  
 \let\W=\Omega
\let\la=\label \let\ci=\cite 
  
\def\nn{\nonumber} \def\bd{\begin{document}} \def\ed{\end{document}}
\def\ds{\documentstyle} \let\fr=\frac \let\bl=\bigl \let\br=\bigr
\let\Br=\Bigr \let\Bl=\Bigl
\let\bm=\bibitem
\let\na=\nabla
\def\tU{{\widetilde U}}
\let\pa=\partial \let\ov=\overline
\def\ie{{\it i.e.\ }}
\newcommand{\be}{\begin{equation}}
\newcommand{\ee}{\end{equation}}
\def\ba{\begin{array}}
\def\ea{\end{array}}
\def\ft#1#2{{\textstyle{{\scriptstyle #1}\over {\scriptstyle #2}}}}
\def\fft#1#2{{#1 \over #2}}
\def\F#1#2{{ F_{#1}^{(#2)} }}
\def\cF#1#2{{ {\cal F}_{#1}^{(#2)} }}

\def\={\, =\, }
\def\+{\, +\, }
\def\-{\, -\, }

\def\R{{\bf R}}
\def\sst#1{{\scriptscriptstyle #1}}
\def\oneone{\rlap 1\mkern4mu{\rm l}}
\def\e7{E_{7(+7)}}
\def\td{\tilde}
\def\wtd{\widetilde}
\def\im{{\rm i}}
\newcommand{\ho}[1]{$\, ^{#1}$}
\newcommand{\hoch}[1]{$\, ^{#1}$}
\newcommand{\bea}{\begin{eqnarray}}
\newcommand{\eea}{\end{eqnarray}}
\newcommand{\ra}{\rightarrow}
\newcommand{\lra}{\longrightarrow}
\newcommand{\Lra}{\Leftrightarrow}
\newcommand{\ap}{\alpha^\prime}
\newcommand{\bp}{\tilde \beta^\prime}
\newcommand{\cB}{{\cal B}}
\newcommand{\cO}{{\cal O}}
\newcommand{\vecx}{\vec{x}}
\newcommand{\vecy}{\vec{y}}
\newcommand{\vecp}{\vec{p}}
\newcommand{\vecq}{\vec{q}}
\newcommand{\tr}{{\rm tr} }
\newcommand{\Tr}{{\rm Tr} }

\newcommand{\cL}{{\cal L}}
\newcommand{\cA}{{\cal A}}
\def\ve{\varepsilon}
\def\vf{\varphi}
\def\F{\Phi}
\def\wg{\wedge}

\def \nn {\nonumber}
\def \rk  {m}
\def \L {{\Lambda}}
\def \ka  { {\kappa }}
\def \S {{ \call S}}

\def\up{\uparrow}
\def\down{\downarrow}

\def \foot {\footnote}
\def \bi{\bibitem}

\def \tr {{\rm tr}}
\def \ha {{1 \over 2}}
\def \td {\tilde}
\def \ci{\cite}
\def \N {{\mathcal N}}
\def \ww {\Omega}
\def \const {{\rm const}}
\def \ss {\sum_{i=1}^3 }
\def \t {\tau}
\def\S{{\mathcal S} }
\def \XX {{\rm X}}

\def \lra {\leftrightarrow}
\def \vom {{\bar \omega}}
\def \E {{\mathcal  E}} \def \J {{\mathcal  J}}
\def \YY {{\rm Y}}

\def \d {\del}
\def \rJ {{J}}
\def \sms {sigma models\ }
\def \sm {sigma model\ }
\def \L {\Lambda}
\def \gl {\ell}
\def \tr {{\rm tr\ }}
\def\z{\zeta}
\def\zi{\zeta_1}
\def\zii{\zeta_2}
\def\K{\mbox{K}}
\def\eE{\mbox{E}}   \def \vt {\vartheta}
\def \vr {\varrho}
\def \wup {w}

\def\dg{\dagger}
\def\a{\alpha}
\def\b{\beta}
\def\e{\varepsilon}
\def\p{\phi}
\def\ap{\alpha^\prime}
\def\I{{\cal I}}

\def\xb{{\bar X}}
\def\Tr{{\rm  Tr}}
\def\tr{{\rm  tr}}

\def \del{\partial}
\def \a {\alpha}
\def \aa {{\a'}}
\def\g{\gamma}
\def\s{\sigma}
\def\z{\zeta}
\def\zi{\zeta_1}
\def\zii{\zeta_2}
\def\ov{\over}

\def\I{{\cal I}}
\def\J{{\mathcal J}}
\def \ok {{1\ov \k}}
\def\LL{{\mathcal L }}
\def \jL {{J}}
\def \om {\omega}
\def \cL {{\mathcal L}} \def \cH {{\mathcal H}}
\def\E{{\mathcal E}}
\def\w{\omega}
\def\b{\beta}
\def\l{\lambda}
\def\eps{\epsilon}
\def\vep{\varepsilon}
\def \De {{\mathcal D}}

\def  \Jt {  {J}_{\rm tot}    }

\def \k {\kappa}
\def\foot{\footnote}
\def \four{{\textstyle {1\ov 4}}}
 \def \third { \textstyle {1\ov 3
}}
\def\det{\hbox{det}}
\def \ci {\cite}

\def \foot {\footnote}
\def \bi{\bibitem}

\def \tr {{\rm tr}}
\def \ha {{1 \over 2}}
\def \tid {\tilde}
\def \vv {{\rm v}}
\def \tl {{\tilde \l}}
\def \XX {{\rm X}}
\def \ta {{\tilde \a}}
\def \fo { {1\ov 4}}
\def \ep {\epsilon}
\def \inti {{\int^{2\pi}_0 {d \sigma \ov 2 \pi}}}

\def \d {\partial}
\def \K {{\rm S}}
\def \el {\ell}
\def \Tr {{\rm Tr}}
\def \P {\Phi}
\def \l  {\lambda}
\def \tl {{\tilde \l}}
\def \bl {{\tilde \l}}
\def \const {{\rm const}}
\def \V {v}

\def \bv {v^*}
\def \vv {{\rm v}}
\def \LL {{\mathcal L}}
\newcommand{\PV}[1]{P_{\!\!_{V_{#1}}}}

\def \S {{\rm S}}
\def \vn {\vec n}
\def \tl {\td \l}
\def \td {\tilde}
\def \Prod {\Pi}
\def \O {{\mathcal O}}
\def \Q {{\rm  Q}}
\def \D {\Delta}
\def \N {{\mathcal N}}
\def\tN{{\tilde N}}

\def \m {\mu}
\def \vs {\vec \s}
\def \ie {i.e.}

\def \cD {{\cal D}}

\def  \le  {\l_{\rm eff}}

\def \rS {{\rm S}}
\def\as{{\a}}
\newcommand{\bra}[1]{\mbox{$\langle #1 |$}}
\newcommand{\ket}[1]{\mbox{$| #1 \rangle$}}

\def\e{\epsilon}

\def \bi{\bibitem}
\def \la {\label}

\def \l {\lambda}
\def\foot{\footnote}
\def \tl  {{\tilde \l}}
\def \sql {{\sqrt \l}}
\def \adss {$AdS_5 \times S^5$\ }
\newcommand{\rf}[1]{(\ref{#1})}
\def \ov {\over}

\def\th{\theta}
\def\Th{\Theta}
\def\vth{\vartheta}
\def\btheta{{\bar\theta}}
\def\ttheta{{{\tilde\theta}}}
\def\bttheta{{{\bar\ttheta}}}
\def\vth{\vartheta}

\def\ra{\rightarrow}
\def\N{{\cal N}}
\def\F{{\cal F}}
\def\uM{\underline{M}}
\def\uN{\underline{N}}
\def\uP{\underline{P}}
\def\cc{\circ}
\def\eqv{\equiv}

\def\ni{\noindent}
\def \ha{{1\ov 2}}
\def \bw {{\rm w}}

\def\r{{\rm r}}
\def \rg{{\rm g}}

\def \J {\mathcal{J}}
\def \del {\partial}

\def\dF{\dot{F}}
\def\dG{\dot{G}}
\def\df{\dot{f}}
\def \E {{\cal E}}

\def \S {{\cal S}}
\def \J {{\cal J}}

\def\ms{\mathcal{S}}
\def\mj{\mathcal{J}}
\def\soj{\fr{\ms}{\mj}}
\def \R {{\bf R}}
\def \om {\omega}
\def \tH {\widetilde H}
\def \bE {\bar E}
\def \x {{\cal X}}

 \def \bb {\bar \beta}
\def \W {{\cal E}}

\def \bi{\bibitem}
\def \la {\label}

\def \mm {{\rm m }}
\def \l {\lambda}
\def\foot{\footnote}
\def \tl  {{\tilde \l}}
\def \sql {{\sqrt \l}}
\def \sqtl {{\sqrt {\tilde \l}}}

\def \HH {{\rm E}}

\def \adss {$AdS_5 \times S^5$\ }

\def \D {\Delta}
\def \thet {\theta}
 \def \t {\tau}
 \def \p {\phi}
 \def \r {\rho}
 \def \rN {{\rm N}}
 \def\tw{{\tilde w}}
 \def\hJ{{J}}
 \def\hw{{w}}
 \def\hl{{\lambda}}
 \def\hth{{\theta}}
 \def\NN{{\cal N}}
 \def \bv {{ \bar w}}
\def \vn {{\vec n}}

\def \ov {\over}

\def \varpi {{\rm w}}
\def \OO {{\cal O}}

\def \rt {{\rm t}}
\def \no {\nonumber }

 \def \bJ {{\rm J}}

\def \adss {$AdS_5 \times S^5$\ }

\def \tp {\tilde \p}

\def \x {\xi}
\def \N {{\cal N}}
\def \sms {sigma-models}
\def \sm {sigma model }
\def \bb {{\bar  \beta}} \def \fo {{ 1 \ov 4}}

 \def \gmn {G_{\m\n}}
 \def \D {{\rm D}}
 \def \bgb {\bb}
 \def \gb {\b}
\def \d {\delta}
\def \fo {{\textstyle {1 \ov 4}} }
\def \tb {\tilde \b}
\def \nD  {\nabla}

\def \S {{\cal S}}

\begin{document}
\overfullrule=0pt
\parskip=2pt
\parindent=12pt
\headheight=0in \headsep=0in \topmargin=0in \oddsidemargin=0in

\vspace{ -3cm} \thispagestyle{empty} \vspace{-1cm}
\begin{flushright} Imperial-TP-AT-6-7\\
\end{flushright}
\begin{center}
 \vspace{2cm}
{\Large\bf On sigma model RG flow, ``central charge'' action \\
\vspace{0.3cm}
 and Perelman's   entropy }
\vspace{0.3cm}

 \vspace{1.5cm} 
A.A.
 Tseytlin\footnote{Also at
 Lebedev  Institute, Moscow.\ \
 tseytlin@imperial.ac.uk
 }

\vskip 0.08cm
 Blackett Laboratory, Imperial College,
London SW7 2AZ, U.K.

\end{center}

 \begin{abstract}
Zamolodchikov's c-theorem  type argument  (and also string theory  effective
action  constructions)
imply that the  RG flow  in 2d sigma model  should be a  gradient one to
all loop orders.
However, the monotonicity  of the flow of the target-space metric  is
not 
obvious since the metric on  the space of metric-dilaton couplings is
indefinite.
To leading (one-loop) order  when the RG flow is  simply the Ricci flow
the monotonicity was proved by Perelman (math.dg/0211159)  by constructing
an ``entropy'' functional  which is essentially the metric-dilaton
action  extremised
with respect to the dilaton with a condition that the   target-space
volume is fixed.
We discuss how to generalize the Perelman's construction to all loop  orders
(i.e. all orders in $\a'$). 
The resulting    ``entropy''    is equal to minus the central charge
at the fixed points,
in agreement with the general claim   of the c-theorem.

\end{abstract}
\newpage

\renewcommand{\theequation}{1.\arabic{equation}}
 \setcounter{equation}{0}

\section{Introduction  }

2d sigma models   containing infinite number of couplings
parametrized by target-space metric tensor $G_{\m\n}(x)$   have  many
interesting connections
to  various problems in
 physics and mathematics.  In particular, they  play an important role
in string theory, describing string
propagation in curved space. In order  to define the 
quantum  stress tensor of the sigma model, i.e. its renormalization on a curved 2d space,
  one is led to the 
introduction an extra scalar coupling function $\p(x)$ or the dilaton.
 The corresponding  metric and dilaton RG  ``beta-functions'' $\b^i$
are then proportional to string effective equations of motion (for
reviews see, e.g., \ci{tse89,calt}).
Assuming that the c-theorem  claim \ci{zam} should apply to the sigma
model 
(with compact euclidean target space) one  should  the existence
of a  (local, covariant)
functional of the metric such that (i) its  gradient is   proportional to
$\b^i$ (with  certain  diffeomorphism terms added), (ii) it 
decreases along  the RG flow and (iii) it  is equal to the central charge
at  fixed points.
While the gradient  property of the flow is relatively easy to establish,
its monotonicity  is much less obvious.  In particular, the direct
generalization of  Zamoldchikov's
proof of the c-theorem \ci{tsec} leads to the ``central charge''
action  which vanishes
at any fixed point, instead of decreasing along the flow.

One may expect that the statement of the c-theorem should apply
provided one considers 
only the  $G_{\m\n}$
metric flow (the flow of the
dilaton  plays,  in a sense, a secondary role).
However, the technical details of the Zamolodchikov-type  construction
 \ci{zam,tsec} of the
``central charge'' functional do not directly apply if one ignores the
dependence on the dilaton.
This suggests that  one needs an alternative  
way of constructing  the corresponding
RG entropy.
The idea of such construction was suggested by Perelman \ci{per}  (see
also a review and generalizations in  \ci{wool})
on the example of the Ricci flow   which is  the 1-loop
approximation to the full
sigma model RG flow. He,  in turn,  was
inspired by the structure of the leading terms in the metric-dilaton
effective action
which first appeared in the string-theory context \ci{ft,cal}.

Below we shall  present a generalization of the Perelman's
construction to all orders in
sigma model loop expansion, i.e.  suggest 
how to prove the c-theorem  for the sigma model. 
We shall first review (in sect.2)
 the basic facts about the structure of sigma model ``beta-functions''
and  the associated
 ``central charge'' action.
In sect.3   we shall define a modification
 of such action  which is equal to minus
the Perelman's
 ``$\l$-entropy''  and interpret the latter as a Lagrange multiplier
for the fixed
volume  condition. We shall argue that this entropy should
grow along the metric RG flow
and is equal to minus  the central charge at the fixed points.

\renewcommand{\theequation}{2.\arabic{equation}}
 \setcounter{equation}{0}

\section{ Review of sigma model results:
Weyl anomaly coefficients and  ``central charge'' action   }

Let us start with a review  of some known facts about
renormalization of  \sm  on  curved 2d space.
One reason to consider quantum \sm on a curved space  is to be able to define 
its stress tensor and its correlators  which enter the standard proof of the c-theorem.
Another is that to be able to define global quantities  that may provide one with a
``c-function''  one needs to consider a 2-space of a topology of a sphere 
(to regularize, in particular, the IR divergences). 
The same   problem  also naturally arises
in the context of  the Polyakov's  approach to critical  string theory
when one considers propagation of a string in $\D$-dimensional
curved target space  with metric $\gmn$ \ci{love,ft,calt}.
 The corresponding action
$I= {1\over { 4 \pi \a' }} \int d^2 z \sqrt {g} \ g^{ab}\    \del_a x^\m  \del_a
 x^\n G_{\m \n}(x) $
 is classically invariant under Weyl rescalings
of the 2d metric $g_{ab}$, implying the decoupling of the conformal
factor of this
metric. Consistency of the critical
string   theory (where   conformal factor of the 2d metric is not a
dynamical field)
 then translates into the requirement of
the  cancellation of the Weyl anomaly.
The renormalizability of  the above  \sm on \ curved 2d background
requires introduction
of  one extra ``hidden'' scalar (dilaton) coupling  which has the same
 dimension as the metric  term
\ci{ft}
\be
 I= {1\over { 4 \pi \a' }} \int d^2 z \sqrt {g} \ [  g^{ab} \del_a x^\m
\del_b  x^\n G_{\m \n}(x)\  +  \ \a' R^{(2)}   \p (x) \ ] \  ,  \la{dil}
\ee
where
$R^{(2)}$ is the curvature of $g_{ab}$.

The two couplings run with renormalization scale according to the corresponding
beta-functions\foot{In our present
notation (opposite to that of \ci{tse89}) the RG evolution toward the  IR
corresponds to $t=-\ln \mu \to + \infty$ ($\mu$ is a momentum  renormalization
scale).}
\be \la{be}
 { d \vp^i \ov dt} = - \b ^i \ , \ \ \ \ \ \ \ \ \
 \vp^i = {(G_{\m\n},\p)}  \ .
\ee
The operator form of the Weyl anomaly relation for the trace of the 2d stress tensor 
is then       \be 2 \pi \a' T^a_a = [\del_a x^\m
\del^a x^\n \bb^G_{\m \n}(x)]\  +  \ \a' R^{(2)} [ \bb^\p (x)] \ , 
\ee
where the $\bb^i$ are the Weyl-anomaly coefficients
that differ from $\b^i$ by certain  diffeomorfism terms  \ci{tse87,shor}
\be
{{\bar \b}^G}_{\mu \nu } =
\b^G_{\mu \nu}  + \nD_ \mu M_\nu   + \nD_ \nu M_\mu      \ ,
\ \ \ \ \ \ \
 \bb^{\phi} = \b^\phi +   M^{\mu} \del_{\mu}\phi   \ ,
 \la{bet}
 \ee
 \be\la{me}
 M_\mu = \a' \del_\m \phi + W_\mu (G)  \ . \ee
  $W_\mu$ is a specific covariant vector constructed out of
 curvature and its covariant derivatives only
 (it is determined from the matrix that governs mixing under
 renormalization of dimension 2 operators).

In dimensional regularization with minimal subtraction one finds to 2-loop
order \ci{honer,frid,ft,tse86,tse87}\foot{The 3-loop $\a'^3$ and 4-loop $\a'^4$
corrections to beta-functions were computed, respectively,
 in \ci{grah} and \ci{jack89}.}
\be \la{on}
{ \b}^G_{\m\n} = \a'  R_{\m\n}  + \ha \a'^2 R_{\m\l\r\s}R_\n^{\
\l\r\s} + O(\a'^3 R^3)
\ee
\be \la{tw}
 { \b}^\p =- \gamma (G) \phi + \omega (G) =
 c_0  - \ha \a'  \nD^2 \p
+ { 1\ov 16}\a'^2 R_{\l\m\n\r} R^{\l\m\n\r} + O(\a'^3 R^3)  \ ,    \ee
where in critical bosonic string
$c_0= {1\ov 6} (\D-26)$ ($\D$ being the total number of coordinates $x^\m$ and $-26$
stands for the measure or ghost  contribution \ci{polk}).
Here $\omega$ is a scalar function of  the curvature and its covariant
 derivatives.
 $\gamma (G)$ is a differential operator (scalar anomalous dimension)
which in the minimal subtraction scheme
has the following general form \ci{frid,calg,tse87}
\be \la{ga}
 \gamma =  \Omega_2^{\m\n} \nD_{(\m} \nD_{\n)}
 + \sum^\infty_{n=3} \Omega_n^{\m_1 ... \m_n}  \nD_{(\m_1} ...
\nD_{\m_n)}  \ ,  \ee
\be \la{kl}
 \Omega_2^{\m\n}  = \ha \a' G^{\m\n}  +
p_1 \a'^3 R^\m_{\a\b\g} R^{\n\a\b\g}  + ... \ , \ \  \ee
\be
\Omega_3^{\m\n\r}  =  q_1 \a'^4 D_\a R_{\b \ \ \g}^{ \ \m\n}
R^{\a\b\g\r}  + ... \ , \ \
\Omega_4^{\m\n\r\l }=  s_1 \a'^4 R^{\m\a\b\n} R^{\r\ \ \ \l}_{\ \a\b}
+ ... \ .  \ee
In the minimal subtraction scheme  $M_\m$ gets a
non-zero contribution  only
 starting with 3 loops
 \be \la{jk}  W_\m = t_1  \a'^3 \del_\m( R_{\l\n\g\r}R^{\l\n\g\r})  +
... \ . \ee
The 3-loop coefficients here are  $p_1 = {3\ov 16}, \ t_1= {1\ov 32} $
\ci{jack88}
 and the 4-loop coefficients
 $q_1,s_1, ...$ were found in \ci{jack89,jack90}.

In general, the Weyl anomaly
coefficients  ${{\bar \b}^G}_{\mu \nu }$  and $\bgb^{\phi}$ satisfy  $\D$
differential identities which can be derived from the condition of
non-renormalisation of the trace of the energy-momentum tensor of the sigma
model \ci{cp,tse87}
\be \la{hi}
 \del_\mu \bgb^{\phi}   -  (  \d^{\l}_\m   \nD^\r  \phi    +  V_\mu^{\l \r})
 {{\bar \gb}^G}_{\l \r  }  =0  \ , \la{cpp} \ee
where the differential operator $V_\mu^{\l \r} $ depends only on
$G_{\m\n}$. To lowest order $V_\mu^{\l \r}{{\bar \b}^G}_{\l \r  }
= \ha \nD^\nu ( {{\bar \b}^G}_{\mu \nu } -
\ha  G_{\m\n} G^{\l \r}{{\bar \b}^G}_{\lambda \r} )  + O(\a'^2) $.
Eq. \rf{cpp}  implies  that  once the metric conformal invariance  equation is imposed, 
$\bb^G_{\m\n} =0$,  then $\bb^\p=$const, and thus $\bb^\p=0$ gives  just
one algebraic equation.

The existence of the
indentity  \rf{cpp}  implying  $ \D$ conditions between      $\ha \D (\D+1) +1$
functions  $\bb^G_{\m\n} $ and $\bb^\p$  would have a natural explanation
if $\bb^G$ and $\bb^\p$  could be obtained  by variation  from a
covariant action  functional  $S(G,\p)$   \ci{zam,polk,tse86}
\be \la{hj}
 { \delta S \ov \delta \vp^i} =  \k_{ij} \bb^j \ , \ \ \ \ \ \ \ \ \ \
\bb^i = \k^{ij}    { \delta S \ov \delta \vp^j}  \ ,  \ee
where $\k_{ij}$ is a non-degenerate covariant operator.
Indeed, the  diffeomorphism invariance of $S$
 implies  the identity
\be \la{kkk}
 \nD_\mu {\d S \over \d G_{\m\n}} -  \ha  {\d S \over \d \p } \nD^\nu \p =0 \
\ 
\ee
which would then relate $\bb^G$ and $\bb^\p$.

Indeed, such action  functional  is easy to find  to leading order in $\a'$
\be \la{act}
 S = \int d^{\D}x \sqrt {G} \ {e}^{- 2
\phi} \bigg[c_0   - \a' \big(  \fo    R  +  \del_\m  \p \del^\m  \p       \big)
 +    O(\a'^2)\bigg] \ .   \ee
Then
\bea \la{lea}
&&{ \bb}^G_{\m\n} = \a' ( R_{\m\n}  + 2 \nD_\m \nD_\n \p) + O( \a'^2) \ ,\\
&&
\bb^\p = c_0 -  \a' (  \ha  \nD^2 \p   + \nD^\m \p  \nD_\m \p) + O( \a'^2)
\la{le}\eea
follow from this action  if
 $\k$ in \rf{hj} is 
\bea \la{kaa}
 \k^{ij}  = { 1 \ov  \sqrt {G} \ {e}^{- 2
\phi} } \left(\begin{array}{cc}
 4 G_{\m \l} G_{\nu \r}  &  G_{\m \n} \\
  G_{\l \r} &    \fo(\D -2) 
\end{array}\right) + O(\a')  \, ,
\eea
\bea \la{koaa}
 \k_{ij}  = {   \sqrt {G} \ {e}^{- 2
\phi} } \left(\begin{array}{cc}
 \fo ( G^{\m \l} G^{\nu \r}  - \ha G^{\m \n} G^{\l \r} )   & \ha  G^{\m \n} \\
  \ha G^{\l \r} &     -2
\end{array}\right) + O(\a')  \, .
\eea
The Lagrangian in \rf{act} can be written as (up to a total derivative)
\be \la{tib}
\tb^\p  \equiv \bb^\p - \fo G^{\m\n} \bb^G_{\m\n} =
c_0   - \fo \a' \big(  R   + 4 \nD^2 \p  - 4 \del_\m  \p \del^\m  \p       \big)
 +    O(\a'^2) \ . \ee
 This  combination \ci{tse87}
 may be interpreted as a ``generalized central charge'' function:
 it appears as the leading term in the
expectation value of the
 trace of the stress tensor ($ 2 \pi  < T^a_a > =  
  \tb^\p     R^{(2)}  + ... $)
  and  is equal to the central charge  at the conformal point where $\bb^G=0$
 (then $\tb^\p=\bb^\p=$const).

 Ref. \ci{tsec}
 put  forward  an  argument
 (based on the idea of the proof of the c-theorem in  \ci{zam})
 that  the ``central charge''  action\foot{In the second equality
below we used  the
 reparametrization invariance of the action and the fact that $\bb^i$
differ from $\b^i$ by
 diffeomorphism terms. Note also that when acting
 on a diffeomorphism-invariant functional one has
 $ \bb^i \cdot { \d \ov \d \vp^i} = \b^i \cdot { \d \ov \d \vp^i}$;
this relation will be used below.}
 \be \la{acti}
 S = \int d^{\D}x \sqrt {G} \ {e}^{- 2
\phi}  \ \tb^\p ( G, \p)  =  \int d^{\D}x \sqrt {G} \ {e}^{- 2
\phi}  \  (\b^\p - \fo  G^{\m\n} \b^G_{\m\n}  )  \      \ee
should have its equations of motion equivalent to $\bb^G=0, \ \bb^\p =0$
to all orders in $\a'$  (provided one chooses an appropriate scheme, i.e.
modulo a  local redefinition of $G_{\m\n}$ and $\p$).
This was indeed confirmed by the explicit sigma model  computations
up to  and including the 4-loop  ($\a'^4$) order \ci{jack88,jack89,jack90}.

This  action has  a remarkable structure. In particular, it  can be rewritten as
\be \la{ct}
 S = - \ha  ( \b^\p  \cdot    { \d   \ov \d \p  }  +  \b^G  \cdot    {
\d  \ov \d G_{\m\n}  }) \ 
 \int d^{\D}x \sqrt {G} \ {e}^{- 2\p}  \  . \ee
Then assuming that $G_{\m\n}$ and $\p$ depend on the renormalization point
and using \rf{be} one finds   \ci{tsec}
\be \la{cut}
 S =  \ha  { d V\ov dt}   \ , \ \ \ \ \ \ \ \
 V \equiv \int d^{\D}x \sqrt {G} \ {e}^{- 2\p}  \ ,  \ee
 i.e. that the ``central charge'' action evaluated on the RG  running couplings
 is simply  the  RG ``time'' derivative of the generalized volume.

More generally,  based on  detailed study of 
renormalization of the sigma model ref.\ci{osb88,osb}
 have constructed an action  that reproduces the Weyl anomaly coefficients, i.e.
 satisfies \rf{hj}, to any order in $\a'$  in an arbitrary 
 (covariant) renormalization scheme. This action hs the following  structure 
\be\la{hio} S_1= \int d^{\D}x \sqrt {G} \ {e}^{- 2
\phi} \big( J_i \bb^i + \r_{ij} \bb^i \bb^j \big)\ , 
  \ee
 where $J_\p=1  +  ..., \ J_G^{\m\n}= - \fo G^{\m\n}+ ...,$ etc.
 are  functions 
 of $\vp_i=(\p, G_{\m\n})$ determined in terms of the renormalization
group quantities.
 The matrix $\r_{ij}$ is,  in principle,  arbitrary (eq.\rf{hj} 
 is satisfied for any $\r_{ij}$) but for a specific
choice of it one can show \ci{osb88} that \rf{hio} becomes (cf. \rf{ct}) 
 \be\la{hiok}
 S_1=     - \ha  ( \b^\p  \cdot    { \d   \ov \d \p  }  +  \b^G  \cdot    {
\d  \ov \d G_{\m\n}  }) \
  \int d^{\D}x \sqrt {G} \ {e}^{- 2 \phi}  \ J_\p (G)  \ ,
  \ee
  where in the  minimal subtraction scheme
  \be\la{kpi}
   J_\p=1 - \fo \a'^2 R_{\m\n\l\r}R^{\m\n\l\r} +  O(\a'^3)\ . \ee
 As was pointed out   in sect. 6 of \ci{tse89}, if one further
redefines the dilaton by  $- \ha \ln J_\p$   the action 
\rf{hiok}
 reduces  to  the simpler-looking
action \rf{acti} or \rf{ct}.\foot{One needs to note that 
under a coupling redefinition $ \b^i\cdot  { \delta \ov \delta \vp^i} = 
\b'^i\cdot  { \delta \ov \delta \vp'^i}$.}
 Thus, 
 there should always exist a scheme in
 which the gradient of  \rf{acti} reproduces $\bb^i$ to all $\a'$ orders.\foot{A possible drawback of this
general argument is that the corresponding scheme choice is somewhat  implicit.
In the dimensional regularization with minimal subtraction the two
actions begin to differ starting with $\a'^3$ order, i.e. to relate
them at 3- and higher loop orders
one needs a certain  field redefinition.}
Refs. \ci{osb88} and  \ci{jack90} found 
 explicitly  the renormalization scheme in which the
action \rf{hio}  of   \ci{osb88}  reduces to the  action \rf{acti} of \ci{tsec} 
at orders  $\a'^3$ and  $\a'^4$ respectively.

\bigskip

Let us note that the representation \rf{cut} is closely related to the
interpretation of the action
whose extrema are equivalent to the vanishing of
 the sigma model  Weyl anomaly coefficients  as the string 
 low-energy effective action for the  graviton and dilaton  modes.
The  effective action can be reconstructed from the string S-matrix.
Its  relation to the sigma model is explained  by the observation  \ci{ft}
that the generating functional
for the string scattering amplitudes can be interpreted as
a partition function  $Z= \int [dx] e^{i I}$ on the 2-sphere
 with the string action  being the \sm
 with couplings which are the string space-time
fields.
The realization   that a  renormalisation of the \sm corresponds to  a
subtraction
of massless poles in the string scattering amplitudes
and that a subtraction of the Mobius volume infinities
can be done by differentiating over the logarithm of the 2d cutoff
led to the expression  for  the  tree-level   closed string
effective action $S$
in terms of the  ``RG time''  derivative of the renormalised
 \sm partition function $Z$  \ci{tsez}:
\be \la{za}
S= { \del Z \ov \del t } =  \b^i \cdot {\del Z \ov \del \vp^i}  \ .
 \ee
Here we  used the RG equation for $Z$: ${ d Z \ov d t }=   
{ \del Z \ov \del t } -  \b^i \cdot {\del Z \ov \del \vp^i}  =0$.
Finally, there exists a  scheme choice  in which  the renormalized
value of $Z$  is simply proportional to the generalized volume $V$ in \rf{cut}
\ci{tsez,tse89}.
This explains the  equivalence between the sigma model RG -motivated
 ``central charge''  action \rf{acti},\rf{cut}
and the  string partition function -motivated  effective action  \rf{za}.

\renewcommand{\theequation}{3.\arabic{equation}}
 \setcounter{equation}{0}

\section{Monotonicity of  RG flow for  $G_{\m\n}$ coupling }

The existence  of the action \rf{acti},\rf{hiok}  whose derivative 
 is proportional \rf{hj}
to the Weyl-anomaly coefficients  implies,  in particular,  that the
RG flow of the metric $G_{\m\n}$  and the dilaton $\p$ couplings
of the \sm \rf{dil} (with a specific choice \rf{bet},\rf{me} of the
diffeomorphism terms)
is a ``gradient  flow''.
This flow is not, however,  monotonic. Indeed, the ``central charge''
action \rf{acti}
vanishes at the fixed points; also, 
$ { d S \ov dt} = -\k_{ij} \bb^i \bb^j $ does not have a definite sign 
since  the ``metric'' $\k_{ij}$
in \rf{hj},\rf{kaa}
is not sign-definite.

At the same time, one may expect (in view of the Zamolodchikov's theorem
\ci{zam} for unitary 2d theories with finite number of couplings)\foot{The proof 
of the  Zamolodchikov's theorem does
not directly apply to the case of  sigma models.
If one tries to repeat the
construction of \ci{zam}
of  the ``central charge function''  (whose gradient
is the beta function)  based on correlators of stress tensor  \ci{tsec}
one needs to introduce the running dilaton coupling and then the
metric on the space of couplings is indefinite (and also the 
value of the central charge function at the fixed point is zero).}
that if one restricts attention  just to the RG flow of the metric $G_{\m\n}$
(and considers the case of compact euclidean signature space)
there should exists an action functional $\S(G)$
 whose gradient is proportional to $\b^G$ and
which  decreases  along the $G_{\m \n}$ flow toward the IR.

A natural guess is that   $\S(G)$  should be closely
related to the functional $S(G,\p)$, e.g., it could 
   be found  by solving for the dilaton, i.e.
by extremising $S(G,\p)$ in $\p$.
The ``secondary'' role  of the dilaton coupling is suggested  by the existence of the 
identity  \rf{cpp}
that expresses its beta-function in terms of the metric beta-function.

The  idea of finding  $\S(G)$   by eliminating   $\p$
from $S(G,\p)$ does not, however,  work directly. At the leading
order  in $\a'$
the combination  $\tb^\p$ in \rf{tib}  has a remarkable property
that its variation over $\p$ with  the measure factor  in \rf{acti} is zero.
The variation of \rf{acti} over $\p$  gives simply
\be
\la{sim}
{ \d S \ov \d \p} =   - 2 \sqrt{G} \ {e}^{- 2\phi} \tb^\p=0\ , 
\ee
i.e.  $\tb^\p=0$, and then the  action $S$
vanishes  even before imposing $\bb^G=0$. The same  property of
$\tb^\p$ is true at least to order
$\a'^4$ and  should be true in general in an appropriate
scheme.\foot{The relation \rf{sim} is valid, in particular, if there
is a scheme in which the dependence
 of $\tb^\p$ on $\p$ in \rf{tib}  is not modified by $\a'$ corrections
to all orders;
 this is actually true to $\a'^3$ order but
 may seem to be  in conflict with the $\a'$-dependence  of the
 operator $\g$ in  \rf{tw},\rf{ga}. But the corresponding  terms  can be
 further redefined away (or integrated by parts) 
  at the level of the action.}

To get a non-trivial functional $\S(G)$ Perelman \ci{per}
suggested  to minimize $S(G,\p)$ in $\p$ while restricting $\p$ to
satisfy the unit volume  condition:\foot{For an earlier closely related suggestion 
in  specific $\D=2$ case see 
\ci{fat}.}
\be
V=\int d^\D x \sqrt{G} \ {e}^{- 2\phi} =1 \ .  \la{vo} \ee
Imposing this condition
 may  be in a  sense 
interpreted as  extremizing $S$ over the constant part of $\tb^\p$
or the central
charge  parameter $c_0$ in \rf{tib}. Indeed,  adding the  constraint \rf{vo}
 to
the action \rf{acti} with the Lagrange multiplier $\l$ we get
the following functional
\be \la{fu}
\hat S =  \int d^{\D}x \sqrt {G} \ {e}^{- 2\phi}\ \tb^\p + \ \
\l \ (\int d^{\D}x \sqrt {G} \ {e}^{- 2\phi}\  -1) \ ,
\ee
i.e.  $\hat S = S({c_0 \to c_0 + \l})    - \l $.

Let us mention in passing   another relation between
actions with unit volume condition and without it. Starting with $S(G,\p)$
one may formally split the dilaton into  constant and non-constant parts
 as follows:
$\p(x) = \p_0 + \tp(x) ,  \ \
\int d^\D x  \sqrt{G} \ {e}^{- 2\tp} = 1  ,$ so that
$
V \equiv \int d^\D x  \sqrt{G} \ {e}^{- 2\phi} = e^{- 2 \p_0}$.
Then $
S(G,\p) = e^{- 2 \p_0} S(G, \tp) = V S(G, \tp) $ and
$\hat S $ in \rf{fu} in which the dilaton is constrained by the volume condition
can be written as (cf. \rf{cut})
\be \la{vv}
\hat S (G,\p) =  S(G, \tp) = { \int d^\D x  \sqrt{G} \ {e}^{- 2\phi}\
\tb^\p (G, \p)\ov
 \int d^\D x  \sqrt{G} \ {e}^{- 2\phi} } = \ha  { d \ov dt} \ln V     \ .  \ee
We shall not use this representation here.

Extremising $\hat S$  with respect to  $\p$ we then get (assuming that
\rf{sim} is true  to all orders in $\a'$)
\be \la{ii}
\tb^\p + \l =0 \ , \ \ee
so that after solving for $\p$, i.e. imposing \rf{ii},  we get
\be \la{jo}
\hat S = \tb^\p = - \l  \ . \ee
Thus $\l$ has an  interpretation of minus the effective central charge.

To leading order in $\a'$  the action $\hat S$ \rf{fu}
and
eq.\rf{ii}  can be written as follows
(see \rf{tib})
\be \la{leab}
\hat S = -\l +    \int d^{\D}x \sqrt {G} \ \Phi
\bigg[ \l + c_0  - \a' ( - \nD^2  + \fo R )   \bigg] \Phi + O(\a'^2)  \ ,
 \ee
 \be \la{dif}
\bigg[  ( - \nD^2 + \fo R ) + O(\a') \bigg]   \Phi =\a'^{-1} (c_0 + \l) \Phi \ , \ee
where
\be  \la{fgg}
\Phi\equiv e^{-\p} \ , \ \ \ \ \ \ \ \ \    \int d^{\D}x \sqrt {G} \
\Phi^2 =1 \ .
 \ee
 The existence of a  solution of this equation with  $\Phi\equiv e^{-\p} > 0$
 requires that  $c_0 + \l$ is
 the minimal eigenvalue of the operator\foot{Let us note in passing  that   the  conformal scalar
 operator in $\D$ dimensions is $- \nD^2  + { \D-2 \ov 4 (\D-1)} R $
 so  that $- \nD^2 + \fo R$ is conformal in the limit $\D\to \infty$.}
  $- \nD^2 + \fo R$
 which always exists
 on a compact space
 \ci{per} (see also \ci{fat}).
 The corresponding eigenfunction will have  no zeros  and can be chosen positive 
 which is what is required for the 
identification of it with
 $e^{-\p}$ (or the inverse of the  effective ``string coupling constant''
  $g_s = e^\p$).
 Thus extremizing $\hat S$  in $\p$ translates (for $c_0=0$) 
 into  choosing  $\l$
 as a minimal eigenvalue of the above Laplacian. 

 To leading order in $\a'$
 when the RG flow defined by  \rf{le},\rf{lea} is simply the Ricci flow the
 Perelman's definition of the functional whose gradient  is $\bb^G$
 (i.e. $\b^G$ with an appropriate diffeomorphism term)
 and which grows monotonically with the  RG flow toward IR ($t \to  \infty$)
 is  simply the minimal eigenvalue $\l$ of \rf{dif}
   (for $c_0=0$). Thus,  in view of \rf{jo},
 \be
 \S(G) \equiv   \hat S(G,\p(G)) =  -  \l(G)   \ . \ee
 Let us    extend this definition to all orders in $\a'$.
 First, the
 variation  of $\S(G) $
 over $G_{\m\n}$ is the same as the $G_{\m\n}$
 variation of $\hat S(G,\p)$ or $S(G,\p)$
 with $\p$ independent of $G_{\m\n}$:  the variation over $\p$ vanishes
 as a consequence of \rf{ii}.
 Then \rf{hj} implies that $\d \S \ov \d G_{\m\n}$ is proportional to
$\bb^G_{\m\n}$.
 In addition, we may ignore the variation of $\sqrt G$ since its
coefficient vanishes
 on the equation for $\phi$.  Then
 \be \la{kka}
 \bb^G_{\m\n} =   \k_{\m\n,\r\s} {\d \S \ov \d G_{\r\s}} \ ,\  \ \ \ \ \ \
 \k_{\m\n,\r\s} =
  {4 G_{\m \r} G_{\n\s} \ov \sqrt G \ e^{-2\p} }
   + O(\a'^2) \ . \ee
Thus $\S$ is a gradient function for the metric RG flow.

 To study
 the monotonicity property of $\S$ we note that
 \be \la{iop}
 { d \ov dt } \S =- \b^G_{\m\n} \cdot {\d \S \ov \d G_{\m\n} }= -
 \bb^G_{\m\n} \cdot {\d \S \ov \d G_{\m\n} }
 = - \bb^G_{\m\n}  \cdot \k^{\m\n,\r\s} \cdot \bb^G_{\r\s} \ . \ee
 Here $\k^{\m\n,\r\s}$   is the inverse  of $\k_{\m\n,\r\s}$
 in   \rf{kka}: on the equation of motion for $\p$ one need not worry 
 about the contribution of the variation of $\sqrt G$ term in the action
 and thus there is no extra term proportional to $- \ha G_{\m\n}
G_{\r\s}$  (cf. \rf{koaa})
 \be \la{jop}
 \k^{\m\n,\r\s} =
  {\fo \sqrt G \ e^{-2\p}  \ G^{\m \r} G^{\n\s}   }
   + O(\a'^2) \ . \ee
 The positivity of $\k^{\m\n,\r\s}$ at leading order in $\a'$
 implies that $\S$ monotonically
 decreases toward the IR ($t\to \infty$) as required of an  effective central
charge \ci{zam},
 while $\l$ grows like an entropy \ci{per}.\foot{
To leading order  in $\a'$
 the relation between Perelman's entropy and the central charge
 was already pointed out in \ci{wool}. The same is true also in
the presence of the $B_{\m\n}$  coupling  \ci{wool}.
Using the monotonicity of $\l$ one
is then able to prove the absence of periodic
 RG trajectories \ci{per,wool}.}
The positivity of $\k^{\m\n,\r\s}$ is obvious 
in perturbation
theory in $\a'$, i.e.
 in sigma model loop expansion.\foot{To  become negative
$\k^{\m\n,\r\s}$ should go through zero but that would require $\a' R
\sim 1$,   invalidating
 perturbation theory expansion.}
 It may be possible to prove it rigorously to all orders
  using the general 
 properties  of renormalization of the sigma model on a curved 
 background as discussed in \ci{osb88,osb}.

 Since the dependence on the dilaton of the  beta-functions in 
 \rf{on},\rf{tw}   is simple
(linear) the same simplicity  should apply to the
effective action. Starting with the action in the 
special scheme \rf{acti}  we shall assume that to all orders in $\a'$ it can be 
put into the form  similar to 
 the leading-order action  \rf{leab} (here we set
$c_0=0$)
 \be \la{leay}
\hat S = -\l +    \int d^{\D}x \sqrt {G} \ \Phi (\l  - \a'  \Delta)  \Phi \
, \ \ \ \ \
\ \ \ \  \ \   \Phi= e^{-\phi} \ , \ee
where 
\be 
\Delta = -  \nD^2  +  U(G)   \ , \ 
\ee
 \be U = {\textstyle { 1 \ov 4}}   R + {\textstyle  { 1\ov 16} }
\a'  R_{\m\n\l\r} R^{\m\n\l\r} +
 {\textstyle { 1\ov 16}} \a'^2 (   R_{\m\n\l\r} R^{\m\n\a\b}
R_{\a\b}^{\ \ \l\r} -
 {\textstyle  {4 \ov 3}  } R_{\k\l\m\n} R^{\k\a\m\b}  R_{\ \a\ \b}^{\l \ \n}) +
 O(\a'^3) . \ee
 For $R^3$ terms we used the result of \ci{mt,jack88}.
 We  have  assumed that  there should exist a  scheme choice
 in which all higher-derivative terms which may be 
  present in the anomalous dimension
 operator $\g$  in \rf{tw} can be integrated by parts in the action  \rf{acti}
 so that $\Delta$ remains  a canonical second-derivative 
 scalar Laplacian as it was at  the leading
order in $\a'$ in  \rf{leab}. This is  indeed what happens  to order $\a'^4$ 
 as was  explicitly  verified in \ci{jack88,jack89}.
  
 The potential function $U(G)$ is a smooth generalization of the
leading-order term
 $\fo R$.  Then  for compact euclidean-signature 
 space the operator $\Delta$ is again positive 
 and its spectrum  should be bounded  from below.
 Then the eigen-function $\Phi$  corresponding to its lowest eigenvalue $\l/\a'$ 
 can  again be  chosen  positive, i.e. there should  exist
 a non-singular solution for $\phi$.\foot{We are grateful to S.
  Cherkis for a clarifying discussion of this point.}
 Combined  with the positivity of the metric  in \rf{jop}, 
  $\l$ will then provide
 the generalization  of the Perelman's entropy to all orders in $\a'$,
 implying the irreversibility of the exact RG flow of the sigma model.

 As we have seen above in \rf{jo}, $\l$ has  also a   meaning of minus the
effective central charge $\tb^\p$, which,  at the fixed point
 $\bar \b^G=0$,  
 is equal to the usual   central charge.
 This is
   in  agreement with the general claim 
    of the c-theorem.
 The construction of the sigma model ``c-function'' {\it a la}  ref.\ci{zam} 
 (i.e. in terms of 2-point functions of stress-tensor components)
 did  lead  \ci{tsec} to $\tb^\p$, but as we explained above
  following Perelman's idea,   to show that the RG flow of $G_{\m\n}$
is  monotonic  one is also  to solve for the dilaton 
and  restrict  its constant part  by the volume
condition \rf{vo}. This  then confirms the validity 
  of the c-theorem for the $G_{\m\n}$ RG flow of the 
  2d sigma model (at least to order $\a'^4$).  

To make   this proof of the c-theorem   rigorous (i.e. to extend it beyond $\a'^4$ order)
  one is to justify   our main assumption  that
  the exact  action  \rf{acti}  can be put  into the form 
 \rf{leay}. This  may  be possible to achieve  using the identities 
 like  \rf{hi}
 following from the  renormalization 
 properties of  composite operators of the sigma model \ci{osb88,osb}.\foot{One may
   also try to  support the ``naturalness''  of the form of 
 \rf{leay} by the following heuristic 
 string-theory argument. Since the $\l$-independent part of \rf{fu},\rf{leay}
 is essentially the tree-level string effective action for the metric and dilaton
 and since $e^{\phi}=\Phi^{-1} $ is the   string coupling ``constant'',  
 the  scaling of the action as $\Phi^2 = e^{-2 \phi}$ is the  direct consequence 
 of the structure of the  string perturbation theory (i.e. of 
 the fact that the Euler number of the 2-sphere  is 2). In general,  the Lagrangian in \rf{leay} may 
 contain also additional terms with derivatives of $\phi$, i.e. depending on 
 $\del_\mu  (\ln \Phi)$. However, such terms ``non-analytic''  in the string coupling 
  would seem  ``unnatural'' 
  -- one could hope  that replacing the string coupling by a 
  function of coordinates should not change the structure of the 
   dependence of the tree-level effective action  on it.}

\bigskip








\bigskip
\bigskip

\section*{Acknowledgments }

We are    grateful to  G. Huisken and T. Oliynyk
 for the invitation
to the Workshop on geometric
 and renormalization group flows at
Max Planck Institute in  Potsdam in November 2006 and for creating
a stimulating  atmosphere  during  the workshop.
We thank S. Cherkis and S. Shatashvili for useful remarks  and discussions. 
 We also acknowledge the support of
PPARC, EU-RTN network MRTN-CT-2004-005104
 and  INTAS 03-51-6346 grants,    and the RS Wolfson award.


\begin{thebibliography}{20}
\bibitem{per}
 G.~Perelman,
 ``The entropy formula for the Ricci flow and its geometric applications,''
 math.dg/0211159.
\end{thebibliography}
\end{document}